\documentclass[11pt, A4]{article}
\usepackage{cite}
\usepackage{graphicx}
\usepackage{epstopdf}
\usepackage{amsmath}
\usepackage{amssymb}
\usepackage{amsfonts}
\usepackage{amsthm}
\usepackage{mathrsfs}

\usepackage[usenames]{color}
\definecolor{AV}{rgb}{0.65,0.0,0}
\usepackage[
      colorlinks=true,
      linkcolor=blue,
      urlcolor=blue,
      filecolor=blue,
      citecolor=red,
      pdfstartview=FitV,
      pdftitle={},
      pdfauthor={},
      pdfsubject={},
      pdfkeywords={},
      pdfpagemode=None,
      bookmarksopen=true
]{hyperref}

\newcommand{\ep}{\epsilon}
\newcommand{\p}{\partial}
\newcommand{\al}{\alpha}
\newcommand{\nn}{\nonumber}
\newcommand{\barphi}{\bar\phi}

\def\m#1{\mathcal#1}
\def\N{{\cal N}}

\begin{document}
\hfill
{\tt ULB-TH/10-36}
\vspace{2cm}

\begin{center}
{\bf \Large Maximally Supersymmetric Yang-Mills in five dimensions in light-cone superspace}
\end{center}
\vspace{1cm}

\begin{center}
{\bf Lars Brink}  \\[2mm]
{\it Department of Fundamental Physics, Chalmers University of Technology,\\ S-412 96 G\"{o}teborg, Sweden}\\[2mm]
{\tt lars.brink@chalmers.se}\\[7mm]

{\bf Sung-Soo Kim}  \\[2mm]
{\it Physique Th\'{e}orique et Math\'{e}matique,  
Universit\'{e} Libre de Bruxelles\\ and\\ International Solvay Institutes,\\
ULB-C.P. 231, B-1050 Bruxelles,
Belgium}\\[2mm]
{\tt sungsoo.kim@ulb.ac.be}\\[15mm]
\end{center}


\begin{abstract}
We formulate maximally supersymmetric Yang-Mills theory in five dimensions in light-cone superspace. The light-cone Hamiltonian is of the quadratic form and the theory can be understood as an oxidation of the $\m N=4$ Super Yang-Mills Theory in four dimensions. We specifically study three-point counterterms and show how these  counterterms vanish on-shell. This study is a preliminary to set up the technique in order to study possible four-point counterterms.
\end{abstract}
\newpage
\section{Introduction}\label{sec:Introduction}

The maximally supersymmetric field theories have been shown to have unique quantum properties. The fact that the $\N=4$ Super Yang-Mills (SYM) theory in $d=4$ is finite in perturbation theory has been known for a long time \cite{Brink:1982wv, Mandelstam:1982cb}. It was a surprising result but the fact that the coupling constant is dimensionless made it possible. The $\N=8$ Supergravity theory in  $d=4$ has on the contrary a dimensionful coupling constant, and it has been believed from the beginning that its perturbation theory should diverge at some loop order. This was the status some thirty years ago and the issue became more academic when the Superstring theory soon afterwards was argued to be a finite quantum gravity theory. However, in recent times, the breakdown in the perturbation expansion for the $\N=8$ Supergravity theory has been questioned and in a very impressive series of papers Bern et al. \cite{Bern:2007hh,Bern:2009kd} have shown that the S-matrix elements are indeed finite up to and including the fourth order loops. They have also shown that the amplitudes satisfy generalized KLT relations \cite{Kawai:1985xq, Bern:2008qj}, in that the amplitudes are essentially the square of the corresponding S-matrix elements for the  $\N=4$ Yang-Mills theory. Can this fact also be valid at higher loops?

This is a very nagging question. There have been a number of papers over the years arguing that possible counterterms could appear at some loop order \cite{Howe:1980th, Kallosh:1980fi, Howe:1983jm, Bossard:2009mn, Bossard:2009sy, Drummond:2010fp}. Presently the conjecture is at the seventh loop order \cite{Kallosh:2008ru, Green:2010sp, Beisert:2010jx, Kallosh:2010mk}. If it is indeed finite it must mean there are structures in these theories that we do not understand. We believe that we understand all the symmetries of the theory, but there could be further symmetries that we have so far not discovered. It could also be some other kind of algebraic structure at work that we have not fully comprehended.  In light-cone superspace, the $\N=8$ theory and the $\N=4$ theory share many formal similarities: both are described by a chiral constrained superfield, and the light-cone Hamiltonians are expressed in terms of a quadratic form based on algebraic relations between the dynamical supersymmetry and the Hamiltonian \cite{Ananth:2005zd, Brink:2008qc}. We will discuss the quadratic form later. It is clear that we have not fully understood its dynamical consequence. At any rate, it is important to try to settle the question about counterterms or possible finiteness and to attack it with different formalisms. Here we will use the light-cone superspace formulation to describe the issue in an alternative formulation. 

In order to show how the light-cone superspace formulation can be helpful we will study a simpler theory in this paper, the $\N=4$ Yang-Mills theory in $d=5$. This theory has a dimensionful coupling constant and it has been argued \cite{Bjornsson:2010wm} that the four-point function first diverges in $d=4 + 6/L$ dimensions (where  $L$ is the loop number). This formula is supposed to continue so that at five loops it should diverge in $26/5$ dimensions and at 6 loops in 5 dimensions. So six loops is the first five-dimensional divergence that is expected. This theory is also interesting since it is a dimensional reduction of the elusive $d=6$ $(2,0)$ theory. It has been argued \cite{Douglas:2010iu} that some divergences in $d=5$ maximally supersymmetric Yang-Mills are difficult to make sense of in the related $d=6$ $(2,0)$ theory.

In this paper we set up the maximally supersymmetric Yang-Mills Theory in $d=5$ in light-cone superspace requiring closure of the SuperPoincar\'e algebra. In superspace, the basic ingredients are the dynamical supersymmetries and we work out the algebra with the dynamical supersymmetries in great detail. We  find that the quadratic form of the Hamiltonian agrees with  the ``oxidation" technique introduced in \cite{Ananth:2004es}.
We then look for possible generalizations of the three-point coupling. Since we now have a dimensionful coupling constant we will show that the SuperPoincar\'e algebra does allow for an infinity of such terms with an ever increasing order of derivatives. However, we will argue that all of those terms can be eliminated by a field redefinition at the level of the equations of motion. The fact that there are no three-point counterterms for this theory is an old result and we do it here in detail to show how our formalism works for these questions. We find that the crucial generator to study counterterms is the dynamical supersymmetry generator and the non-existence of such terms for this generator implies non-existence of possible counterterms for the Lagrangian too.
\setcounter{equation}{0}
\section{Lightcone formulation in five dimensions}
\subsection{Notation: Symplectic spinors}
In five dimensions, 
there are three transverse directions $x^1,x^2,x^3$. We denote the coordinates by\begin{align}
x&=\frac{1}{\sqrt2}(x^1+ix^2),& \bar x&=\frac{1}{\sqrt2}(x^1-ix^2),&x^3,\cr
\bar\p&=\frac{1}{\sqrt2}(\p_1-i\p_2),& \p&=\frac{1}{\sqrt2}(\p_1+i\p_2),&\p_3,
\end{align}
and the light-cone coordinates by 
\begin{align}
x^{\pm}&=\frac{1}{\sqrt2} (x^0\pm x^4 )\, ;& \p^{\mp}&=\frac{1}{\sqrt2}(-\p_0\mp\p_4)= - \p_{\pm},
\end{align}
so that 
\begin{align}
\p^-x^+&=-1=\p^+x^-,& \p \bar x &= 1= \bar\p x= \p_3 x^3.
\end{align}
We choose $x^+$ as an evolution parameter (the light-cone time). Without loss of generality, one can set $x^+=0$.

In order to formulate the maximally supersymmetric Yang-Mills theory in a superspace whose $R$-symmetry is $SO(5)_R \approx Sp(4)_R$ and the little group is $SO(3)\approx SU(2)$, we introduce a symplectic Grassmann variable $\theta^{i\al}$ such that
\begin{align}
\bar \theta_{i\al}=\theta^{j\beta}C_{ji}\epsilon_{\beta\al},
\end{align}
where $\al, \beta,\ldots=1,2$ are the $SU(2)_{spin}$ indices and the $SU(2)$ invariant tensor $\epsilon^{\al\beta}$ satisfies
\begin{align}
\epsilon^{\al \beta} &= \overline{\epsilon_{\al\beta}},&&\epsilon^{\al\beta}\epsilon_{\beta \gamma}=-\delta^{\al}{}_\gamma,&&(\ep^{12}=\ep_{12}=1),
\end{align}
and the $SO(5)$ spinor indices are labelled by $i,j,\ldots=1,2,3,4$. 
The charge conjugation $C_{jk}$ matrix\footnote{One can also think of it as the invariant tensor of $Sp(4)$}
is antisymmetric, $C^{jk} = -{C^{kj}}$, and satisfies 
\begin{align}
C^{jk} = \overline{C_{jk}},&&(C^{ij})^\dagger &= -C^{ij}&& C^{jk}C_{kl}=-\delta^{j}{}_l.
\end{align}
It follows that the consistency condition is easily checked
\begin{align}
\theta^{i\al}=\overline{\bar \theta_{i\al}}= \overline{(\theta^{j\beta})} C^{ji}\epsilon^{\beta\alpha}= \theta^{k\gamma}C_{kj}\epsilon_{\gamma\beta}C^{ji}\epsilon^{\beta\al}=\theta^{i\al}.
\end{align}
The commutation relations among $\theta^{i\al}$ and their derivatives are given by
\begin{align}
\{\theta^{i\al},\theta^{j\beta}\}&=0&& \left\{\frac{\p}{\theta^{i\al}},\frac{\p}{\theta^{j\beta}}\right\}=0&&
\left\{\theta^{i\al},\frac{\p}{\theta^{j\beta}}\right\}=\delta^{i}{}_j\delta^{\al}{}_\beta.
\end{align}
The kinematical supersymmetries, the spectrum generating parts of supersymmetry, are represented as 
\begin{align}
q^{i\al}&= -\frac{\p}{\p \bar \theta_{i\alpha}}+\frac{i}{\sqrt2}\theta^{i\al}\p^+,\cr
\bar q_{i\al}&=
\frac{\p}{\p \theta^{i\al}}-\frac{i}{\sqrt2}\bar \theta_{i\alpha}\p^+
=-q^{j\beta}C_{ji}\epsilon_{\beta\al},\label{twoindexsusy}
\end{align}
and satisfy
\begin{align}
\{q^{i\al},\bar q_{j\beta}\} &= i\sqrt2\p^+\delta^{i}{}_{j}\,\delta^{\al}{}_{\beta},
\end{align}
or equivalently
\begin{align}
\{q^{i\al},q^{j\beta}\} &= -i\sqrt2\p^+C^{ij}\,\ep^{\al\beta}.
\end{align}

\subsection{Little group}
The $SU(2)$ little group generators are easily defined by introducing
\begin{align}
x^{\al\beta}=x^{\beta\al}&:= \frac{1}{\sqrt2}\ep^{\al\gamma}\left( x\cdot\sigma\right)_\gamma{}^\beta
\end{align}
where $\ep^{\al\gamma}= i(\sigma^2)^{\al\gamma}$. In terms of matrix form, it reads
\begin{align}
\begin{pmatrix}
x^{11}	&	x^{12}\\
x^{21}	&	x^{22}
\end{pmatrix}
=
\begin{pmatrix}
x	&	-\frac{1}{\sqrt2}x^{3}\\
-\frac{1}{\sqrt2}x^{3}	&	-\bar x
\end{pmatrix}
.
\end{align}
The corresponding transverse derivatives are then given by
\begin{align}
\p_{\al\beta}=\p_{\beta\alpha} :=-\frac{1}{\sqrt2}\left( \p\cdot\sigma\right)_\alpha{}^\gamma \,\ep_{\gamma\beta}= \begin{pmatrix}
\bar \p	&	-\frac{1}{\sqrt2}\p_3\\
-\frac{1}{\sqrt2}\p_3		&	- \p
\end{pmatrix},
\end{align}
such that 
\begin{align}
[\,\p_{\al\beta}\,,x^{\gamma\rho}\,] 
=\delta^{\gamma}{}_{(\al}\delta^{\rho}{}_{\beta)},
\end{align}
or
\begin{align}
\p_{11}x^{11}=1,\qquad \p_{22}x^{22}=1,\qquad \p_{12}x^{12}=\frac12.
\end{align}
The indices are raised and lowered by $\ep^{\al\beta}$ and $\ep_{\al\beta}$, and thus 
\begin{align}
\p^{\al\beta}=\ep^{\al\gamma}\ep^{\beta\rho}\p_{\gamma\rho}= 
\begin{pmatrix}
-\p	&	\frac{1}{\sqrt2}\p_3\\
\frac{1}{\sqrt2}\p_3		&	\bar\p
\end{pmatrix}
.
\end{align}


The orbital angular momenta are represented by
\begin{align}
L_1
&= \left[ (x^{11}+x^{22})\p_{12}+x^{12}(\p_{11}+\p_{22})\right],\cr
L_2
&=i \left[ x^{12}(\p_{11}-\p_{22})- (x^{11}-x^{22})\p_{12}\right],\cr
L_3
&=\left(x^{11}\p_{11}-x^{22}\p_{22} \right)= x\bar\p-\bar x \p.\label{L}
\end{align}
The raising and lower operators
\begin{align}
L_{+} &= L_1+iL_2
=\sqrt2 (x^3\p -x\p_3),\cr
L_{-} &= L_1-iL_2
=-\sqrt2 (x^3\bar\p -\bar x\p_3),\label{L+L-}
\end{align}
satisfy ${L_+}^\dagger=L_-$ and 
\begin{align}
[L_+,L_-] &= 2 L_3,& [L_3, L_+]& =+L_+,&  [L_3, L_-] &=-L_-.
\end{align}

The $SU(2)_{spin}$ generators are constructed in terms of the kinematical supersymmetry generators \eqref{twoindexsusy} by contracting the $SO(5)$ indices 
\begin{align}
S^\al{}_\beta 
 = \frac{1}{2i\sqrt2\p^+}\left(
q^{i\al}\bar q_{i\beta} - \frac12 \delta^\al{}_\beta q^{i\gamma}\bar q_{i\gamma}\right)\ ,
\end{align}
which obey
\begin{align}
[ S^\al{}_\beta, S^\gamma{}_\rho ] = \delta^\gamma{}_\beta S^\al{}_\rho - \delta^\al{}_\rho S^\gamma{}_\beta\  .
\end{align}
The raising and lowering operators are given by
\begin{align}
S_+ =& ~S^1{}_{2}=\frac{1}{2i\sqrt{2} \,\p^+}q^{i1}\bar q_{i2}
,\\
S_- =&~ S^2{}_{1}=\frac{1}{2i\sqrt{2} \,\p^+}q^{i2}\bar q_{i1}
,\\
S_3 =&~  \frac{1}{2}(S^1{}_{1}-S^2{}_{2}) =\frac{1}{4i\sqrt{2} \,\p^+}(q^{i1}\bar q_{i1}-q^{i2}\bar q_{i2})
,\label{S3-1}
\end{align}
satisfying ${S_+}^\dagger=S_-$ as well as  the $SU(2)$ commutation relations
\begin{align}\label{su2al}
[S_+,S_-] &= 2 S_3,& [S_3, S_+]& =+S_+,&  [S_3, S_-] &=-S_-.
\end{align}
Together with \eqref{L} and \eqref{L+L-}, one can form the full $SU(2)$ little group generators as
\begin{align}
M_{+} 
&=
\sqrt2 (x^3\p -x\p_3)+S_+,\cr
M_{-} &=
-\sqrt2 (x^3\bar\p -\bar x\p_3)
+S_-,\\
M_3 &=
x\bar\p-\bar x \p +S_3.\nn
\end{align}

In a similar way, the $SO(5)_R\approx Sp(4)_R$ R-symmetry generators $R^{ij}$ are expressed as quadratic operators 
\begin{align}
R^{ij}=R^{ji}= \frac{1}{i\sqrt2\p^+}q^{i\al}\ep_{\al\beta}q^{j\beta},
\end{align}
which obey
\begin{align}
[R^{ij},R^{kl}] &= C^{jk}R^{il}+C^{jl}R^{ik}+C^{ik}R^{jl}+C^{jl}R^{ik}.
\end{align}

\subsection{New notation}
Since $\theta^{i2}$ is related to the complex conjugate of $\theta^{i1}$, it suffices to use one kind of Grassmann variable
\begin{align}
\theta^i&\equiv\theta^{i1},&&
\bar \theta_i\equiv \bar \theta_{i1} \,(= C_{ij}  \theta^{j2}),
\end{align}
and thus 
\begin{align}
\Big\{\theta^{i},\frac{\p}{\p\theta^{j}}\Big\}&=\delta^{i}{}_j=
\Big\{\bar \theta_{j},\frac{\p}{\p\bar \theta_{i}}\Big\}.
\end{align}
It follows that the kinematical supersymmetry generators are written as
\begin{align}
q^{i} &
=-\frac{\p}{\p \bar\theta_i}+\frac{i}{\sqrt2}\theta^i\p^+
,& \bar q_{i}&
=\frac{\p}{\p \theta^i}-\frac{i}{\sqrt2}\bar \theta_i\p^+,
\end{align}
satisfying
\begin{align}
\{q^i,\bar q_j\}&=i\sqrt2 \p^+ \delta^i{}_j,
\end{align}
and the $SU(2)_{spin}$ generators are 
\begin{align}
S_+ =&
\frac{1}{2i\sqrt{2} \,\p^+}q^{k} C_{kl}q^{l} ,\cr
S_- =&-\frac{1}{2i\sqrt{2} \,\p^+}\bar q_{k} C^{kl}\bar q_{l} ,\\
S_3 =&~  
\frac{1}{4i\sqrt{2} \,\p^+}(q^{j}\bar q_{j}-\bar q_{j}q^{j}).\nn
\end{align}

The chiral field which captures all the physical degrees of freedom reads
\begin{align}
\phi^a(y) =&~
\frac{1}{\p^+}A^a(y)+\frac{i}{\p^+}\theta^k\bar\chi^a_k(y)+\frac{i}{\sqrt2} \theta^k{C}_{kl}\theta^l\,A^a_3(y) +\frac{i}{\sqrt2}
\theta^k(C\gamma^I)_{kl}\theta^l\, D^a_I(y)\cr
& +\frac{\sqrt2}{6}\theta^i\theta^j\theta^k\epsilon_{ijkl}\chi^{l\,a}(y)
+\frac{1}{12}\epsilon_{ijkl}\theta^i \theta^j \theta^k \theta^l 
\p^+ \bar A^a(y),
\end{align}
where $a$ is the gauge index, $I=1,\ldots5$, and 
the chiral (light-cone) coordinate $y$ is defined as
\begin{align}
y& = (x,\bar x, x_3, x^+, y^-\equiv x^--\frac{i}{\sqrt2}\theta^i\bar\theta_i ).
\end{align}
The component fields representing bosonic degrees of freedom are associated with even powers of the Grassmann variables, while the fields representing fermionic degrees of freedom are associated with odd powers of the Grassmann variables. For instance, the three vector degrees of freedom are denoted by $A^a$, its complex conjugate $\bar A^a$, and real $A^a_3$; the five scalars are by $D^a_I$ which are also real. The eight fermionic degrees of freedom are denoted by $\bar\chi^a_k$  and $\chi^{k\,a}$.

The superfield is (anti-) chiral:
\begin{align}
d^i \phi(y)&=0,& \big(~ \bar d_i \barphi(\bar y)&=0,~\big) 
\end{align}
where the (anti-) chiral derivatives are given by
\begin{align}
d^{i} &
=-\frac{\p}{\p \bar\theta_i}-\frac{i}{\sqrt2}\theta^i\p^+
,&& \left(\bar d_{i}
=\frac{\p}{\p \theta^i}+\frac{i}{\sqrt2}\bar \theta_i\p^+, \right)
\end{align}
satisfying
\begin{align}
\{d^i,\bar d_j\}&=-i\sqrt2 \p^+ \delta^i{}_j,
&&\{d^i, q^j\}=0=\{d^i,\bar q_j\} .
\end{align}
The superfield is subject to the inside-out constraint
\begin{align}
d^id^j\phi^a =\frac{1}{2}\epsilon^{ijkl}\bar d_k\bar d_l \barphi^a.\label{IO}
\end{align}
This inside-out constraint naturally relates $A^a$ and $\bar A^a$ by complex conjugation, and leads to  
\begin{align}
C_{ij}=\frac{1}{2}\epsilon_{ijkl}C^{kl},
\end{align} 
implying that the totally antisymmetric tensor $\epsilon^{ijkl}$ is proportional to
a linear combination of quadratic $C$'s and thus yielding 
\begin{align}
\epsilon_{ijkl}= C_{ij}C_{kl}+C_{jk}C_{il}+C_{ik}C_{jl}.\label{epCC}
\end{align}

All the symmetry generators are then defined on the chiral superfield. There are kinematical and dynamical generators. 
Kinematical generators $Q_{kin}$ act linearly, 
\begin{align}
\delta_{Q_{kin}}\phi^a = Q_{kin}\,\phi^a.
\end{align}
For example, the kinematical supersymmetry transformations are 
\begin{align}
\delta_{q}\phi^a &=q\,\phi^a,&
\delta_{\bar q}\phi^a &=\bar q\,\phi^a.
\end{align}
It is straightforward to see that 
\begin{align}
\begin{pmatrix}
q^i\\
C^{ij}\bar q_j
\end{pmatrix}
\end{align}
forms a doublet under $SU(2)$.
\begin{align}
[\delta_q, \delta_{M_{3}}]\phi^a&=\frac{1}{2} \delta_q\phi^a,&
[\delta_{\bar q}, \delta_{M_{3}}]\phi^a&=-\frac{1}{2} \delta_{\bar q}\phi^a,
\end{align}
where we have used 
\begin{align}
\delta_{S_{3}}\phi = \frac{1}{4i\sqrt{2} \,\p^+}[q^{i},\bar q_{i}]\phi= \left( -1 +\frac{1}{2}\theta^i\bar q_i\right)\phi.
\end{align}
It is instructive to see how the component fields transform under $S_+$. 
First, we observe 
\begin{align}
\delta_{S_+} \phi &= {S_+} \phi  =\frac{i}{\sqrt2}\p^+ \theta{C}\theta
\Big( \frac{1}{\p^+}A + \cdots\Big).\label{S+} 
\end{align}
and notice that the  transformation can also directly act on the component fields
\begin{align}
\delta_{S_+} \phi &= \frac{1}{\p^+}(\delta_{S_+}A) + \cdots.\label{S++}
\end{align}
We then compare \eqref{S++} to \eqref{S+} to obtain
\begin{align}
\delta_{S_+} A &= 0,&\delta_{S_+} A_3 &= A ,&
 \delta_{S_+}\bar A &= -2A_3,\cr
\delta_{S_+}\bar \chi_i &= 0,&
\delta_{S_+} \chi^i &=C^{ij} \bar \chi_j , &
 \delta_{S_+}D_I &= 0.
\end{align}
This shows that $A, A_3, \bar A$ form a triple of $SU(2)$ representing physical degrees of freedom of a gauge field in five dimensions. Five scalars $D_I$ are $SU(2)$ singlet. The fermions form a doublet 
under the $SU(2)$
\begin{align}
\begin{pmatrix}
C^{ij}\bar \chi_j\\
\chi^i
\end{pmatrix}.
\end{align}

\setcounter{equation}{0}
\section{Interacting Hamiltonian} 
\subsection{Dynamical supersymmetry transformations}
To build the free dynamical supersymmetry generators which transform as the $\bf 2$ of $SU(2)$, we 
start with the highest weight $\p$ of the triple made   out of the transverse momenta
\begin{align}
[M_+,\p]=0.
\end{align}
Together with $q^i$, one can construct the highest weight state with spin $3/2$, 
$\p q^i$. We apply $M_-$ on this, to find the state of spin $1/2$
\begin{align}
-\sqrt2 \p_3 q^i + \p C^{ij}\bar q_j.
\end{align}
The highest weight state for the dynamical supersymmetry transformations then can be constructed by switching the relative sign and with a proper normalization factor
\begin{align}
-\frac{1}{\sqrt2} \p_3 q^i - \p C^{ij}\bar q_j.
\end{align}

Yet another simpler way is 
to take the antisymmetric product of  the kinematical supersymmetry generator $q^i$ transforming as  $\bf 2$ under $SU(2)$ and  the transverse derivatives $\p_{\al\beta}$ transforming as $\bf 3$. This leads to the same result
\begin{align}
\ep^{\al\beta}\p_{\beta\gamma}q^{i\gamma} \qquad \longrightarrow\qquad -\frac{1}{\sqrt2} \p_3 q^i - \p C^{ij}\bar q_j.
\end{align}
Therefore, we can introduce the (free) dynamical supersymmetry generators as
\begin{align}
{Q^{i\al}}\phi^a =
-\frac{\ep^{\al\beta}\p_{\beta\gamma}q^{i\gamma}}{\,\p^+}\phi^a.\label{freedyna}
\end{align}
It is convenient to define the free dynamical supersymmetry generators as
\begin{align}
\bar q_{-i}& \equiv \overline{Q}_{i2}= \frac{ \p}{\p^+}\bar q_{i}\phi^a-\frac{\p_3}{\sqrt2 \p^+}C_{ij}q^j ,
\cr
q^i_- &\equiv Q^{i2}=\frac{ \bar\p}{\p^+}q^{i}\phi^a-\frac{\p_3}{\sqrt2 \p^+}C^{ij}\bar q_j
\end{align}
so that, upon dimensional reduction to $d=4$, they reproduce the same expressions for the dynamical supersymmetry transformations in four dimensions. These generators satisfy the supersymmetry commutation relation
\begin{align}
\{q^i_- ,\bar q_{-j}\} =i\sqrt2 \delta^i{}_j \frac{1}{\p^+}\Big(\p\bar\p+\frac{\p_3^2}{2}\Big).
\end{align}

For the interaction part, the same analysis can be applied as in \cite{Ananth:2005zd}. It is then straightforward to see that the interaction part of the dynamical supersymmetry transformations does not depend on the transverse derivatives. This means that the form of the interacting dynamical supersymmetry transformations in $d=5$ is the same as that of $d=4$ Super Yang-Mills. Therefore, the full dynamical supersymmetry transformations for $d=5$ Super Yang-Mills are given by
\begin{align}
\delta_{\varepsilon\bar q_-}\phi^a&= \delta^{free}_{\varepsilon\bar q_-}\phi^a +\delta^{int}_{\varepsilon\bar q_-}\phi^a \cr 
&= \varepsilon^i\bar{q}_{-i}\phi^a -  gf^{abc}\frac{1}{\p^+}(\varepsilon^i \bar q_{i}\phi^b \p^+ \phi^c),\cr
&=\varepsilon^i \left[ \frac{ \p}{\p^+}\bar q_{i}\phi^a-\frac{\p_3}{\sqrt2 \p^+}C_{ij}q^j \phi^a
-  gf^{abc}\frac{1}{\p^+}( \bar q_{i}\phi^b \p^+ \phi^c)
\right],\label{susybarq}
\end{align}
where $g$ is the coupling constant. It follows from the inside-out constraint \eqref{IO} that
\begin{align}
\delta_{\bar\varepsilon q_-}\phi^a &=\bar\varepsilon_i \left[ \frac{ \bar\p}{\p^+} q^{i}\phi^a-\frac{\p_3}{\sqrt2 \p^+}C^{ij}\bar q_j \phi^a
-  gf^{abc}\frac{d^{(4)}}{2\p^{+3}}( q^{i}\barphi^b \p^+ \barphi^c)
\right],\label{susyq}
\end{align}
where $d^{(4)} =\frac{1}{4!}\epsilon_{ijkl}d^id^jd^kd^l$. The full dynamical supersymmetries transform as a highest weight under the  little group $SU(2)$
\begin{align}
[\delta_{\bar q_-}, \delta_{M_{+}}]\phi^a&=0,&
[\delta_{\bar q_-}, \delta_{M_{3}}]\phi^a&=+\frac{1}{2} \delta_{\bar q_-}\phi^a.\label{hw}
\end{align}
Note that in \eqref{susybarq} there cannot be a term of involving 
\begin{align}
C_{ij}q^j\barphi^b\p^+\phi^c + \cdots,
\end{align}
where $\cdots$ refers to the terms ensuring chirality. It is because such a term would lead to a transformation that is not highest weight. This really shows the uniqueness of  the non-linear term. 

It is also worth noting that  the full dynamical supersymmetry transformations can also be written in the form of a \emph{covariant} derivative
\begin{align}
\delta_{\varepsilon \bar q_-}\phi^a =\varepsilon^i\frac{1}{\p^+}\Big[ (\m D^{ab})^{\gamma}\bar q_{i\gamma} \phi^b \Big],
\end{align}
where
\begin{align}
(\m D^{ab})^\gamma=\delta^{ab}\p^{1\gamma}
+gf^{abc}\p^+\phi^c\epsilon^{2\gamma}.
\end{align}
Such covariant derivative structure was already observed in \cite{Ananth:2005zd}.
We note that this suggests that the dynamical supersymmetry transformations for the maximally supersymmetric Yang-Mills in other dimensions may also be written in terms of a covariant derivative. The existence of the covariant derivative reflects the residual light-cone gauge symmetry \cite{Kallosh:2010mk}\footnote{
We thank Pierre Ramond for pointing out this residual gauge symmetry.}.

The Hamiltonian transformation on the superfield $\delta_{\m P^-}\phi$ can be obtained from the supersymmetry algebra
\begin{align}
 [\,\delta_{\bar\varepsilon  q_-}\,,\,\delta_{\varepsilon \bar q_-}\,]\phi^a = \sqrt2 \,\bar\varepsilon\cdot\varepsilon\,\delta_{\m P^-}\phi^a.
\end{align} 
The free Hamiltonian transformation is
\begin{align}
\delta^{free}_{\m P^-}\phi^a 
&=-i\frac{1}{\p^+}\left(\p\bar\p+\frac{\p_3^2}{2}\right)\phi^a,
\end{align}
as expected. The interacting Hamiltonian transformation is also obtained in the same way 
\begin{align}
\delta^{int}_{\m P^-}\phi^a 
&= [\,\delta^{free}_{\bar\varepsilon q_-}\,,\,\delta^{int}_{\varepsilon \bar q_-}\,]\phi^a
+
[\,\delta^{int}_{\bar\varepsilon q_-}\,,\,\delta^{free}_{\varepsilon \bar q_-}\,]\phi^a +
[\,\delta^{int}_{\bar\varepsilon q_-}\,,\,\delta^{int}_{\varepsilon \bar q_-}\,]\phi^a,\label{Pint}
\end{align}
where the interacting dynamical supersymmetries are given in \eqref{susybarq} and \eqref{susyq}. It is tedious but straightforward; one needs to repeatedly use \eqref{epCC} and the inside-out constraints \eqref{IO}. The interacting Hamiltonian transformation to order $g$ is then given by
\begin{align}
\delta^{int}_{\m P^-}\phi^a =&
-i g f^{abc}\left[\frac{1}{\p^+} ( \bar\p \phi^b\p^+\phi^c ) +
\frac{d^{(4)}}{2\p^{+3}}(\partial \bar \phi^b \p^+\barphi^c )
\right]\cr
&+\frac{g}{2}f^{abc}\frac{1}{\p^{+2}}\left[\p^+\p_3\phi^b\bar d C\bar d \phi^c 
+2 C^{rs}\bar d_r\p^+\phi^b\bar d_s\p_3 \phi^c+\frac{\p_3}{\p^+}\bar d C\bar d\phi^b \p^{+2}\phi^c\right]\quad\cr
&+\m O(g^2),\label{deltaintP}
\end{align}
where $\bar d C\bar d=\bar d_r C^{rs}\bar d_s$.

\subsection{Hamiltonian and generalized transverse derivatives}
Super Yang-Mills theories in various dimensions in the light-cone superspace share many similarities.   
 One of the most salient features is that the light-cone Hamiltonian is of  quadratic form. It was first noticed in $d=4,\,\m N=4$ Super Yang-Mills \cite{Ananth:2005zd} and then also confirmed for $d=3, \,\m N=8$ BLG theory \cite{Belyaev:2010kt}. Higher dimensional theory also respect such structure. For example, the Hamiltonian for $d=10,\, \m N=1$ Super Yang-Mills is still of the quadratic form~\cite{Belyaev:2009rj}\footnote{We thank D. Belyaev for informing us the quadratic form of $\m N=1$ Super Yang-Mills in ten dimensions.}.

In \cite{Ananth:2004es} it was discussed that $\m N=4$ Super Yang-Mills in four dimensions can be easily oxidized to higher dimension by replacing the transverse derivatives in the interaction terms with generalized transverse derivatives $\nabla$. The essence of $\nabla$ and $\overline{\nabla}$ is that they are covariant under the Lorentz little group and their forms are quadratic in the (anti-) chiral derivatives. Since the transverse derivatives appear only in the three-point interaction terms, the quartic interaction terms remain unaltered through the oxidation procedure. We show that it is true in five dimensions as well and the form of the generalized transverse derivatives is given by 
\begin{align}
\nabla&= \p+\frac{i}{8} \frac{\p_3}{\p^+}d^kC_{kl}d^l ,\label{nabla}\\
\overline{\nabla}&= \bar\p+\frac{i}{8}\frac{\p_3}{\p^+}\bar d_kC^{kl}\bar d_l.\label{barnabla}
\end{align}

To see this, let us recall the quadratic form of  the light-cone Hamiltonian in \cite{Ananth:2005zd}. The quadratic form basically dictates the fact that the Hamiltonian is the square of the dynamical supersymmetries. In other words, the fully interacting Hamiltonian $H$ can be expressed as a quadratic form of the dynamical supersymmetries   
\begin{align}
H &=\frac{1}{2\sqrt2}(\m W^a_i, \m W^a_i)\equiv\frac{i}{\sqrt2} \int d^{13}z \,\overline{\m W^a_i}\frac{1}{\p^+}\m W^a_i,
\end{align}
where $d^{13}z= d^5 x\,d^4\theta \,d^4\bar\theta$ and $ \varepsilon^i\m W^a_i= \delta_{\varepsilon\bar q_-}\phi^a$
\begin{align}
\m W^a_i&= \frac{ \p}{\p^+}\bar q_{i}\phi^a-\frac{\p_3}{\sqrt2 \p^+}C_{ij}q^j \phi^a
-  gf^{abc}\frac{1}{\p^+}( \bar q_{i}\phi^b \p^+ \phi^c),\cr
\overline{\m W^a_i}&= \frac{ \bar\p}{\p^+}q^{i}\barphi^a-\frac{\p_3}{\sqrt2 \p^+}C^{ij}\bar q_j \barphi^a
-  gf^{abc}\frac{1}{\p^+}( q^{i}\barphi^b \p^+ \barphi^c).
\end{align}
Because of the inside-out constraint \eqref{IO}, the free Hamiltonian  
is written as
\begin{align}
H^{free} &= \frac{i}{\sqrt2}\int d^{13}z \Big(\frac{\bar \p}{\p^+}q^j-\frac{\p_3}{\sqrt2 \p^+}C^{jk}\bar q_k \Big)\barphi^a \frac{1}{\p^+}\Big(\frac{\p}{\p^{+}}\bar q_j-\frac{\p_3}{\sqrt2 \p^+}C_{jl}q^l \Big)\phi^a \cr
&= -\frac{i}{2\sqrt2}\int d^{13}z \barphi^a \Big(\frac{\p\bar \p}{\p^{+3}}+\frac{\p_3}{2\p^{+3}}\Big)\{q^k,\bar q_k\}\phi^a\cr
& = \int d^{13}z\,\barphi^a \Big(\frac{2\p\bar \p}{\p^{+2}}+\frac{\p_3}{\p^{+2}}\Big)\phi^a,
\end{align}
where the cross term involving $\p\p_3$ (or $\bar\p\p_3$) vanishes by itself due to $\{\bar q_j,\bar q_k\}=0$ (or  $\{q^j, q^k\}=0$).
The three-point interacting Hamiltonian can be expressed in a simple form if we use 
two nontrivial identities
\begin{align}
f^{abc} \int d^{13}z \, \frac{\p}{\p^+}{\bar q}_i\phi^a \frac{1}{\partial^{+2}} (q^i
{\bar \phi}^b \partial^+{\bar \phi}^c)
=-\frac{4i\sqrt{2}}{3}f^{abc}\int d^{13}z \,\frac{1}{\partial^+}\phi^a{\bar 
\phi}^b\partial{\bar \phi}^c,
\end{align}
and
\begin{align}
&f^{abc}\int  d^{13}z \,\frac{\p_3}{\p^+}q^i \phi^a C_{ij}\frac{1}{\p^{+2}}(q^j\bar\phi^b\p^+\barphi^c)\cr
&=~\frac{1}{3}f^{abc}\int d^{13}z \,\left(\frac{1}{\p^+}\phi^a\barphi^b \frac{d^iC_{ij}d^j}{\p^+}{\p_3} \barphi^c+
\frac12 \phi^a\barphi^b \frac{d^iC_{ij}d^j}{\p^{+2}}{\p_3} \barphi^c\right).
\end{align}
The detailed proof of these identities is presented in Appendix \ref{app:B}.
The three-point interacting Hamiltonian $H^{(3)}$ is then 
\begin{align}
H^{(3)} 
&=-i\frac{g}{\sqrt2}f^{abc}\int d^{13}z \, \bigg\{\Big(\frac{\p}{\p^+}{\bar q}_i\phi^a -\frac{\p_3}{\sqrt2\p^+}C_{ij}q^j  \phi^a \Big)\frac{1}{\partial^{+2}} (q^i
{\bar \phi}^b \partial^+{\bar \phi}^c)+c.c.\bigg\}\cr
&=-\frac{4}{3}gf^{abc}\int d^{13}z\bigg\{\frac{1}{\partial^+}\phi^a \bar 
\phi^b \Big(
 \p+\frac{i}{8} \frac{\p_3}{\p^+}d^kC_{kl}d^l \Big)
 \barphi^c 
+ \frac{1}{\partial^+}\barphi^a 
\phi^b \Big(
\bar\p+\frac{i}{8}\frac{\p_3}{\p^+}\bar d_kC^{kl}\bar d_l\Big)
\phi^c\cr
&\qquad\qquad \qquad+\frac{i}{16} \phi^a \barphi^b\frac{\p_3}{\p^{+2}}d^kC_{kl}d^l\barphi^c
 +\frac{i}{16} \barphi^a \phi^b\frac{\p_3}{\p^{+2}}\bar d_kC^{kl}\bar d_l\phi^c\bigg\},\end{align}
where the last two terms are canceled each other due to the inside-out constraint and the antisymmetry of $f^{abc}$. For instance, the last term is rewritten as
\begin{align}
f^{abc}\barphi^a \phi^b\frac{\p_3}{\p^{+2}}\bar d_kC^{kl}\bar d_l\phi^c =
-f^{abc}\barphi^b \phi^a\frac{\p_3}{\p^{+2}}d^kC_{kl} d^l\barphi^c
\end{align}
and thus cancels out the third term. Hence, we have
\begin{align}
H^{(3)}=-\frac{4}{3}gf^{abc}\int d^{13}z\,\bigg(\frac{1}{\partial^+}\phi^a
\barphi^b \nabla\barphi^c+\frac{1}{\partial^+}\barphi^a\phi^b \overline{\nabla}\phi^c\bigg),
\end{align}
where $\nabla$ and $\overline{\nabla}$ are given in \eqref{nabla} and \eqref{barnabla}, respectively. 

The four point interacting Hamiltonian does not contain any transverse derivatives and thus its form is the same as for $\m N=4$ SYM in four dimensions. It is, in fact, true in all other dimensions as well. We remark as was done in \cite{Ananth:2005zd}, that it is crucial that $f^{abc}$ must satisfy the Bianchi identity. In order to show that the order $g^2$ terms from the quadratic form is indeed the same as the four point interaction terms this condition follows. This confirms that $f^{abc}$ is indeed a structure constant of Lie algebras.
Combining all terms, we obtain that the full Hamiltonian is 
\begin{align}
H= \int{d^{13}}z\, \m H,
\end{align}
where
\begin{align}
\m H=&~\bar\phi^a_{}\,\frac{2\partial\bar\partial + {\partial_3}{}^2}{\partial^{+2}}\phi^a
-\frac{4}{3}gf^{abc}\Big(\frac{1}{\p^+}\,\barphi^a\phi^b \overline{\nabla} \phi^c+\frac{1}{\p^+}\phi^a\barphi^b\nabla\barphi^c\Big)\cr 
&+g^2f^{abc}f^{ade}\Big( \frac{1}{\partial^+}
(\phi^b\partial^+\phi^c)\frac{1}{\partial^+}
(\bar\phi^d\partial^+\bar\phi^e)
+\frac{1}{2}\phi^b\bar\phi^c\phi^d\bar\phi^e\Big),
\end{align}
which agrees with the result of \cite{Ananth:2004es}. The Lagrangian is then 
\begin{align}
\m L=&-\bar\phi^a_{}\,\frac{\Box}{\partial^{+2}}\phi^a
+\frac{4}{3}gf^{abc}\Big(\frac{1}{\p^+}\,\barphi^a\phi^b \overline{\nabla} \phi^c+\frac{1}{\p^+}\phi^a\barphi^b\nabla\barphi^c\Big)\cr 
&-g^2f^{abc}f^{ade}\Big( \frac{1}{\partial^+}
(\phi^b\partial^+\phi^c)\frac{1}{\partial^+}
(\bar\phi^d\partial^+\bar\phi^e)
+\frac{1}{2}\phi^b\bar\phi^c\phi^d\bar\phi^e\Big).
\end{align}

\setcounter{equation}{0}
\section{Three-point on-shell counterterms}

When the full superPoincar\'e algebra was constructed for the $d=4$ maximally symmetric theory  it was easy to see that the construction was indeed unique. However, for the  $d=5$ theory with a dimensionful coupling constant there could be other non-linear terms in the dynamical generators where higher order in the transverse derivatives are compensated by higher powers of the coupling constant $g$. Such terms could be interpreted as possible counterterms.
In this section, we will hence extend our construction of dynamical supersymmetry transformations to also include such possible terms. 

In supersymmetric theories, the Hamiltonian is not a fundamental quantity. It is rather the dynamical supersymmetry that is fundamental since the Hamiltonian can be obtained from it. Especially in light-cone superspace, the Hamiltonian is of a quadratic form written in terms of the dynamical supersymmetries. If the full superPoincar\'e algebra is unbroken at quantum level, it is then natural to search for possible counterterms in the dynamical supersymmetry transformations. This means that the full dynamical supersymmetry transformations would be split into 
\begin{align}
\delta^{full}_{\bar q-}\phi&= \delta^{cl}_{\bar q-}\phi + \delta^{ct}_{\bar q-}\phi,
\end{align}
where $\delta^{cl}_{\bar q-}\phi$ are dynamical supersymmetry transformations that yield the classical action and $\delta^{ct}_{\bar q-}\phi$ are the terms which account for counterterms via the quadratic form 
\begin{align}
\m H^{full}= \m H^{cl}+ \m H^{ct},
\end{align}
or  equivalently,
\begin{align}
\m L^{full}= \m L^{cl}+ \m L^{ct}.
\end{align}

This may leads to a better understanding of the quantum property of a given theory. As a first attempt, we consider three-point counterterms. To this end, it is useful to introduce the coherent state-like form \cite{Brink:2008qc, Brink:2008hv} which is not only a way to sustain chirality of the transformations but also  the most general expression that commutes with all kinematical symmetry generators. Let us first examine three-point one-loop counterterms. To construct possible one-loop counterterms in the dynamical supersymmetry generator, we introduce one transverse derivative as well as a kinematical supersymmetry generator in an $SU(2)$ invariant way as in \eqref{freedyna}. This means that we may have terms of the form
\begin{align}
g^3 f^{abc} \frac{\ep^{\al\beta}\p_{\beta\gamma}}{\p^+}q^{i\gamma} \phi^b \phi^c,\qquad {\rm or}\qquad
g^3 f^{abc} \frac{\ep^{\al\beta}\p_{\beta\gamma}}{\p^+}q^{i\gamma} \phi^b \barphi^c.\label{1-loop}
\end{align}
By checking \eqref{hw}, however, it is easy to see that such terms cannot be highest weight of $SU(2)$, irrespective of the location of $\p^+$. We note that the same reasoning holds for all odd loop counterterms. To see this, recall that odd loop counterterms in the dynamical supersymmetry transformations require an odd number of transverse derivatives. Then the $SU(2)$ invariance enforces that at least one index in the transverse derivatives $\p^{\alpha\beta}$ must be contracted. This means that the effects of such derivatives over the one in (4.4) are nothing but an $SU(2)$ singlet since $\partial^{\alpha\beta}\partial_{\beta\gamma}=\frac{1}{2}\partial^{\beta\rho}\partial_{\beta\rho} \delta^{\alpha}{}_{\gamma}$. The remaining index structure is then of the same form as in \eqref{1-loop}, and thus not of heightest weight.

On the other hand, even loop three-point counterterms in the dynamical supersymmetry transformation have different characteristics compared to the odd loop counterterms, because they are associated with the even number of transverse derivatives and the $SU(2)$ indices can be contracted among themselves.  For instance, a possible two loop three-point counterterm is comprised of two transverse momenta with the fifth power of the coupling constant and thus it should be of the form
\begin{align}
\delta^{ct}_{\bar q-}\phi^a &\sim g^5f^{abc}\frac{1}{\p^{+(2M+1)}}\big( \bar q_{i}\p^{\al\beta}\p^{+M}\phi^b \p_{\al\beta}\p^{+(M+1)} \phi^c\big)+\cdots,
\end{align}
to be a highest weight under $SU(2)$.  The integer $M$ will not be not determined until we check the commutation relations with other dynamical transformations, e.g, $[\delta^{ct}_{\bar q_-}, \delta_{\m J^-}]\phi=0$. However, the exact value here is not relevant for the discussion below. For simplicity, we choose $M=0$: 
\begin{align}
\delta^{ct}_{\bar q-}\phi^a &\sim g^5f^{abc}\frac{1}{\p^{+}}\big( \bar q_{i}\p^{\al\beta}\phi^b \p_{\al\beta}\p^{+} \phi^c\big)+\cdots.
\end{align}
By requiring that these counterterms must satisfy all the commutation relations with the kinematical Super-Poincar\'e generators, one finds that they should be only of the form
\begin{align}
\delta^{ct}_{\bar q-}\phi^a &\propto g^5f^{abc}\frac{1}{\p^{+}}\Big(
 \p^{\al\beta}\p_{\al\beta}\frac{\bar q_{i}}{2\p^{+2}}\phi^b \p^{+2} \phi^c
- \bar q_{i}\p^{\al\beta}\phi^b \p_{\al\beta}\p^+ \phi^c
+ \bar q_{i}\phi^b \frac{\p^{\al\beta}\p_{\al\beta}}{2}\phi^c\Big), \nn
\end{align}
which can be written as a coherent state-like form
\begin{align}
\delta^{ct}_{\bar q-}\phi^a &=c\,g^5f^{abc}
\left(\frac{\p}{\p w_{\al\beta}}\frac{\p}{\p w^{\al\beta}}\right)
\frac{1}{\p^+}\left( E \bar q_{i} \p^+\phi^b E^{-1}\p^{+2} \phi^c
\right)\Big|_{w^{\alpha\beta}=0},\label{dyn-cc}
\end{align}
where $c$ is a constant and 
\begin{align}
E= \exp\left[
\frac{w^{\al\beta}\p_{\al\beta}}{\p^+}
\right].
\end{align}

We note that the coherent state-like forms are, in fact, closely related to equations of motion. To see this, introduce a combination of two chiral superfields
\begin{align}
A \phi^b B\phi^c,\label{AB}
\end{align}
where $A$ and $B$ are some bosonic/fermionic operators acting on the fields which are not explicitly dependent on the coordinates. Now we take a d'Alembertian on \eqref{AB}
\begin{align}
\Box (A \phi^b B\phi^c) &= (-2\p^+\p^- - \p^{\al\beta}\p_{\al\beta}) (A \phi^b B\phi^c),
\end{align}
where $\p^{\al\beta}\p_{\al\beta}= - 2\p\bar\p -\p^2_3$. Implementing the equations of motion
\begin{align}
\p^-\phi^a = -\frac{\p^{\al\beta}\p_{\al\beta}}{2\p^+}\phi^a+\delta^{int}_{\m P^-}\phi^a,
\end{align}
we find that
\begin{align}
\Box (A \phi^b B\phi^c) = & ~
( \frac{\p^{\al\beta}\p_{\al\beta}}{\p^+}A \phi^b \p^+B\phi^c)
-2 ( \p^{\al\beta}A \phi^b \p_{\al\beta} B\phi^c)\cr
&+( \p^+A\phi^b \frac{\p^{\al\beta}\p_{\al\beta}}{\p^+}B \phi^c )
-2\p^+\delta^{int}_{\m P^-}(A \phi^b B\phi^c).\label{boxonAB}
\end{align}
Notice that the last term in \eqref{boxonAB} is of higher order in the field $\phi^a$. It is then obvious that the terms quadratic in the field $\phi^a$ are exactly of the form of the coherent state-like form above. Hence, 
the equations of motion enable us to express a d'Alambertian operator acting on two chiral fields in terms of a coherent state-like form 
\begin{align}
\Box (A \phi^b B\phi^c) &=  \frac{\p}{\p w_{\al\beta}}\frac{\p}{\p w^{\al\beta}}\left( 
E \p^+A \phi^b\, E^{-1}\p^+B  \phi^c
\right)\Big|_{w=0}
+\m O(g).
\end{align}
This means that using the equations of motion, we can rewrite \eqref{dyn-cc} as
\begin{align}
\delta^{ct}_{\bar q-}\phi^a &=g^5f^{abc}\frac{\Box}{\p^+} \Big(
 \bar q_{i} \phi^b \p^+\phi^c \Big)
 + \m O(g^6),\label{cc2loop}
\end{align}
or by reintroducing $M$,
\begin{align}
\delta^{ct}_{\bar q-}\phi^a &=g^5f^{abc}\frac{\Box}{\p^{+(2M+1)}} \Big(
 \bar q_{i} \p^{+M}\phi^b \p^{+(M+1)}\phi^c \Big)
 + \m O(g^6).
\end{align}

Moreover, the counterterm \eqref{dyn-cc} can be generalized to all even loop orders.
\begin{align}
\delta^{ct}_{\bar q-}\phi^a =&\sum_{l, \, {\rm even}} \,c_l\,g^{1+2l}f^{abc}\cr
&{\left(\frac{\p}{\p w_{\al\beta}}\frac{\p}{\p w^{\al\beta}}\right)}^{l/2}
\frac{1}{\p^{+(2M+1)}}\left( E \bar q_{i} \p^{+(M+1)}\phi^b E^{-1}\p^{+(M+2)} \phi^c
\right)\Big|_{w^{\alpha\beta}=0}.\label{dyn-cc-1}
\end{align}
Checking the full super-Poincar\'e algebra we can see that they are indeed representations of the full algebra and hence possible counterterms. We are, however, here interested in counterterms that survive on the mass shell and hence we check if we can rewrite them as we have done with the two-loop counterterm in \eqref{cc2loop}. Indeed, using the mass shell condition we find that 
\begin{align}
\delta^{ct}_{\bar q-}\phi^a &=\sum_{l, \, {\rm even}} c_l \, g^{1+2l}f^{abc}\frac{{\Box}^{l/2}}{\p^{+(2M+1)}} \Big( \bar q_{i}\p^{+M} \phi^b \p^{+(M+1)}\phi^c
\Big)
 + {\rm higher~point~ functions}
.\label{cc2loop-1}
\end{align}

We now show that these counterterms will lead to counterterms that can be eliminated in the Hamiltonian. Consider the calculation of the Hamiltonian variation $\delta^{int}_{\m P^-}\phi^a $ in \eqref{Pint}. If we introduce $\delta^{ct}_{\bar q-}\phi^a $ from \eqref{cc2loop-1} into \eqref{Pint}, we find 

\begin{equation}
\delta^{ct}_{\m P^-}\phi^a = \sum_{l, \, {\rm even}} c_l \, g^{1+2l}\,{\Box}^{l/2}\, \delta^{\widetilde{int}}_{\m P^-}\phi^a + {\rm higher\,\, point\,\, functions},
\end{equation}
where $\delta^{\widetilde{int}}_{\m P^-}\phi^a$ is a term of a similar structure as \eqref{deltaintP} which has the undetermined power $M$ but no structure constant $g$. Again, the precise form of $\delta^{\widetilde{int}}_{\m P^-}\phi^a$ is irrelevant for our discussion.  
The equations of motion then become
\begin{equation}
\Box\,\phi^a = - 2\delta^{int}_{\m P^-}\partial^+\phi^a -2   \sum_{l, \, {\rm even}} c_l \, g^{1+2l}\,{\Box}^{l/2}\, \partial^+ \delta^{\widetilde{int}}_{\m P^-}\phi^a  +\,{\rm higher\,\, point\,\, functions.}
\end{equation}
We can so make a field redefinition
\begin{equation}
\phi{}'{}^a =  \phi^a -2  \sum_{l, \, {\rm even}} c_l \, g^{1+2l}\,{\Box}^{l/2-1}\, \delta^{\widetilde{int}}_{\m P^-}\partial^+\phi^a.
\end{equation}
to obtain the new equation of motion where we have dropped the prime on the field.
\begin{equation}
\Box\,\phi^a =  -2\partial^+\delta^{int}_{\m P^-}\phi^a  +\,{\rm higher\,\, point\,\, functions.}
\end{equation}
We have hence shown that all the possible three-point counterterms can be eliminated (or rather pushed up to higher point functions) when we use the mass-shell condition. 

We must also check if there could be some other ans\"atze for the dynamical supersymmetry transformation in terms of $\phi^a$. However, as we pointed out before, there is no other starting point that can be of  highest weight under $SU(2)$ without derivatives. The only way to introduce space derivatives is to let the indices saturate each other. Hence the net effect of introducing space derivatives into an expression would not change the overall transformation under $SU(2)$ and we conclude that there are no possible counterterms for the three-point interaction.

\section{Conclusion and discussions}
In this paper, we have constructed the maximally supersymmetric Yang-Mills theory in five dimensions in light-cone superspace by introducing symplectic Majorana spinors.
We found that this theory shares many similarities with $\m N=4$ Super Yang-Mills in four dimensions: The dynamical supersymmetry transformations possess a covariant derivative structure which encodes the full interactions as seen in the $\m N=4$ theory in four dimension. The Hamiltonian is of a quadratic form which is the same as that of $\m N=4$ theory. We also showed that the theory can be easily seen as an oxidation from $d=4$ to $d=5$ via generalized transverse derivatives in the light-cone superspace.

This theory is supposed to diverge at the six-loop order for the four-point function. We have examined here possible three-point counterterms. In this formalism which is equivalent to the light-cone gauge, we found that there are no possible odd loop three-point counterterms. 
For even loops, we showed how all possible three-point counterterms can be absorbed into the kinetic term by implementing the equations of motion and making a field redefinition. 
Our method is a practical hands-on way to study possible counterterms and in a forthcoming paper we will study possible four-point counterterms.
\section*{Acknowledgments}
We thank Dmitry Belyaev, Jonas Bj\"{o}rnsson, Guillaume Bossard, Michael Green, and Pierre Ramond for useful discussions and Michael Douglas for a correspondence. L.B. wants to thank Charles Thorn for a very useful question during a seminar some years ago. S.K. thanks Department of Fundamental Physics, Chalmers University of Technology for hospitality during his visit where most of this work was carried out.
S.K. is supported by IISN - Belgium (conventions 4.4511.06 and 4.4514.08), by the Belgian Federal Science Policy Office through the Interuniversity Attraction Pole P6/11.
\section*{Appendix}
\appendix
\setcounter{equation}{0}
\section{Useful identities}\label{app:B}
\setcounter{equation}{0}
\noindent\underline{\bf Chiral functions and superfields }\\
A chiral function $f(y)$ can be written as 
\begin{align}
f(y) &= e^{\frac{i}{\sqrt2}\theta^m\bar\theta_m\p^+}f(x),
\end{align}
It is then easy to see that 
\begin{align}
d^mf(y) &= 0,&\bar d_mf(y) &= i\sqrt2\,\bar \theta_m\p^+ f(y),\cr
q^m f(y)& = i\sqrt2 \,\theta^m \p^+ f(y)\ , & \bar q_m f(y)& = 0.
\end{align}
By complex conjugation, one easily finds similar relations for an antichiral function.
For superfields, (anti-) chiral condition, $d\phi=0$ ($\bar d\barphi=0$), yields
\begin{align}
\frac{\partial}{\partial\bar\theta}\phi& =-\frac{i}{\sqrt2}\theta\partial^+\phi,&
\frac{\partial}{\partial\theta}\bar\phi& =-\frac{i}{\sqrt2}\bar\theta\partial^+\bar\phi,
\label{theta}
\end{align}
or
\begin{align}
q^m \phi &= i\sqrt2 \theta^m\p^+ \phi,&
\bar q_m \bar\phi &= -i\sqrt2 \bar\theta_m \p^+ \bar\phi.\label{qphi}
\end{align}
The following commutation relations are useful when one finds the form of the interaction terms from the quadratic forms of the light-cone Hamiltonian:
\begin{align}
[ \,\bar d_i, \, 
\theta\frac{\partial}{\partial\theta}+\bar\theta\frac{\partial}{\partial\bar\theta}\,]\,=\,\bar 
q_i , \label{usefulcomm1}
\end{align}
and 
\begin{align}
[\bar d_i ,\theta^kC_{kl}\frac{\p}{\p\bar\theta_l}] = -C_{ij}d^j.\label{usefulcomm2}
\end{align}\\


\noindent\underline{\bf Identity }\\
For a chiral combination $X^c$, 
\begin{align}
\label{id2}
f^{abc}\int\frac{1}{\partial^+{}^2}\bar\phi^a\phi^b X^c=0.
\end{align}
This can be easily seen by implementing the inside-out constraint on $\phi^b$.
In a similar way, one also finds that
\begin{align}
f^{abc}C_{ij}\int \frac{1}{\p^{+2}} q^i \phi^a q^j \barphi^b Y^c = 0,
\end{align}
where $Y^c$ is either chiral or antichiral. Another useful identity involves two chiral derivatives
\begin{align}
f^{abc}\int  (d^i \barphi^a C_{ij} d^j \barphi^b) X^c = 0,
\end{align}
which is due to the antisymmetric property of $C_{ij}$ and $f^{abc}$.\\

\noindent{\underline{\bf First 3-point function identity }}\\
As shown in \cite{Ananth:2005zd}, the identity 
\begin{align}\label{identity}
-f^{abc} \int \frac{\p}{\p^+}{\bar q}_i\phi^a \frac{1}{\partial^{+2}} (q^i
{\bar \phi}^b \partial^+{\bar \phi}^c)
=\frac{4i\sqrt{2}}{3}f^{abc}\int\frac{1}{\partial^+}\phi^a{\bar 
\phi}^b\partial{\bar \phi}^c\
\end{align}
is an important identity that is crucial to see the quadratic form of the light-cone Hamiltonian.  Here we prove it with more details. 

Perform the partial integral with respect to $\bar q_i$ and then use \eqref{qphi} to obtain
\begin{align}
-f^{abc}\int  \frac{\p}{\p^+}{\bar q}_i\phi^a \frac{1}{\partial^{+2}} (q^i
{\bar \phi}^b \partial^+{\bar \phi}^c) 
&= i\sqrt2
f^{abc}\int  \phi^a \frac{\p}{\partial^{+2}} ({\bar \theta}_i\frac{\p}{\p\bar\theta_i}
{\bar \phi}^b \partial^+{\bar \phi}^c).\label{startingpt}
\end{align}
The integrations by parts with respect to $\frac{\p}{\p\bar\theta_i}$ allow us to express \eqref{startingpt} as
\begin{align}
&4i\sqrt2 f^{abc}\int \frac{1}{\p^{+2}}\phi^a\p(\bar \phi^b\p^+\bar\phi^c)\cr
&-i\sqrt{2}f^{abc}\int \frac{1}{\p^{+2}}\bar\theta\frac{\p}{\p\bar\theta}\phi^a\p(\bar\phi^b\p^+\bar\phi^c)
-i\sqrt{2}f^{abc} \int\frac{1}{\p^{+2}}\phi^a\p(\bar\phi^b\bar\theta\frac{\p}{\p\bar\theta}\p^+\bar\phi^c).\label{st2}
\end{align}
Using \eqref{id2}, we see that the first term of \eqref{st2} becomes
\begin{align}
4i\sqrt2 
f^{abc}\int \frac{1}{\partial^+{}^2}\phi^a\partial\bar\phi^b\partial^+\bar\phi^c.
\end{align}
The second term of \eqref{st2} is 
\begin{align}
-i\sqrt{2}f^{abc} \int
\frac{1}{\p^{+2}}\bar\theta\frac{\p}{\p\bar\theta}\phi^a\p\bar\phi^b\p^+\bar\phi^c
-i\sqrt{2}f^{abc} \int \frac{1}{\p^{+2}}\bar\theta\frac{\p}{\p\bar\theta}\phi^a
\bar\phi^b\p^+\p\bar\phi^c.
\label{st3}
\end{align}
The integrations by part with respect to $\p^+$ acting on $\barphi^c$ on the last term of \eqref{st2} yield
\begin{align}
&-i\sqrt{2}f^{abc} \int\frac{1}{\p^{+2}}\phi^a\p(\bar\phi^b\bar\theta\frac{\p}{\p\bar\theta}\p^+\bar\phi^c)\label{st4}\\
&= 
i\sqrt{2}f^{abc} \int\,
\frac{1}{\p^{+}}\phi^a \p\bar\phi^b\bar\theta\frac{\p}{\p\bar\theta}\bar\phi^c 
+
\frac{\p}{\p^{+2}}\phi^a\p\p^+\bar\phi^b\bar\theta\frac{\p}{\p\bar\theta}\bar\phi^c
-\frac{\p}{\p^{+2}}\phi^a\bar\phi^b\bar\theta\frac{\p}{\p\bar\theta}\p\p^+\bar\phi^c.\nn
\end{align}
Combining the second term of \eqref{st3} and the last two terms of \eqref{st4}, we find that
\begin{align}
-i\sqrt2 f^{abc}\int \bar\theta\frac{\p}{\p\bar\theta} \left(\frac{1}{\p^{+2}}\phi^a\barphi^b\p\p^+\barphi^c\right) =0,
\end{align}
thanks to \eqref{id2}.
The remaining terms are then (the integral symbol $\int$ is omitted from here on)
\begin{align}
& 4i\sqrt2 
f^{abc}\frac{1}{\partial^+{}^2}\phi^a\partial\bar\phi^b\partial^+\bar\phi^c
-i\sqrt2f^{abc}\frac{1}{\partial^+{}^2}\bar\theta\frac{\partial}{\partial\bar\theta}\phi^a\partial\bar\phi^b\partial^+\bar\phi^c
-i\sqrt2f^{abc}\bar\theta\frac{\partial}{\partial\bar\theta}\bar\phi^b\frac{1}{\partial^+}\phi^a\partial\bar\phi^c\cr
&\equiv {\rm I+II+III}
\ ,\nn
\end{align}
which we call ${\rm I}$, ${\rm II}$, and ${\rm III}$ respectively.

We now work on the ${\rm I}$ term: the integration by parts with respect to 
$\partial^+$ (acting on $\barphi^c$) yields
\begin{align}
{\rm I}
&=4i\sqrt2 f^{abc}\frac{1}{\partial^+}\phi^a\bar\phi^b\partial\bar\phi^c 
-4i\sqrt2 
\frac{1}{\partial^+{}^2}\phi^a\bar\phi^c\partial^+\partial\bar\phi^b \cr
&=4i\sqrt2 f^{abc}\frac{1}{\partial^+}\phi^a\bar\phi^b\partial\bar\phi^c ,\label{I} 
\end{align}
where we used \eqref{id2} in the last step.

For the ${\rm II}$ term, it follows from\footnote{There is a typo in \cite{Ananth:2005zd}: (B.30) of \cite{Ananth:2005zd} should be of the form \eqref{typo} above.}
\eqref{theta} that
\begin{align}
{\rm II}&=-i\sqrt2f^{abc}\frac{1}{\partial^+{}^2}\frac{\partial}{\partial\bar\theta_i}\phi^a\,(\bar\theta_i\partial^+\bar\phi^b) \,\partial\bar\phi^c
=-i\sqrt2f^{abc}\frac{1}{\partial^{+}}{\theta^i}\phi^a\frac{\p}{\p\theta^i}\bar\phi^b\partial\bar\phi^c\cr
&=-i\sqrt2f^{abc}\theta\frac{\partial}{\partial\theta}\bar\phi^b\frac{1}{\partial^+}\phi^a\partial\bar\phi^c
.\label{typo}
\end{align}
Combining ${\rm II}$ and ${\rm III}$, we obtain
\begin{align}
{\rm II+III}=-i\sqrt2f^{abc}(\theta\frac{\partial}{\partial\theta}+\theta\frac{\bar\partial}{\partial\bar\theta})\bar\phi^b\frac{1}{\partial^+}\phi^a\partial\bar\phi^c.
\end{align}
This can be further simplified, if we use 
\begin{align}
[ \,\bar d_i, \, 
\theta\frac{\partial}{\partial\theta}+\bar\theta\frac{\partial}{\partial\bar\theta}\,]\,=\,\bar 
q_i , \label{usefulcomm1}
\end{align}
and the inside-out relations on $\phi^c$, as 
\begin{align}
{\rm II+III}&=-i\sqrt2f^{abc}\frac{\bar d^4}{2}\left[(\theta\frac{\partial}{\partial\theta}+\bar\theta\frac{\bar\partial}{\partial\bar\theta})\bar\phi^b\frac{1}{\partial^+}\phi^a\right]\frac{\partial}{\p^{+2}}\phi^c\cr
&=-i\sqrt2f^{abc}\left[ \frac{\ep^{ijkl}}{2\cdot3!}
\bar q_i \barphi^b \frac{\bar d_{jkl}}{\p^+}\phi^a
+(\theta\frac{\partial}{\partial\theta}+\bar\theta\frac{\bar\partial}{\partial\bar\theta})\barphi^b\p^+\barphi^a
\right]\frac{\p}{\p^{+2}}\phi^c.
\end{align}
The use of the inside-out constraint and \eqref{qphi} yields
that 
\begin{align}
{\rm II+III}&=-i\sqrt2f^{abc}\left[\frac{i}{\sqrt2} 
\bar q_i \barphi^b d^i\phi^a
+(\theta\frac{\partial}{\partial\theta}+\bar\theta\frac{\bar\partial}{\partial\bar\theta})\barphi^b\p^+\barphi^a
\right]\frac{\p}{\p^{+2}}\phi^c\cr
&=-2i\sqrt2f^{abc}\left[(\theta\frac{\partial}{\partial\theta}+\bar\theta\frac{\bar\partial}{\partial\bar\theta})\barphi^b\p^+\barphi^a
\right]\frac{\p}{\p^{+2}}\phi^c.
\end{align}
Since the first term vanishes by itself due to \eqref{theta} and the antisymmetric property of $f^{abc}$
\begin{align}
-2i\sqrt2 f^{abc}\,\theta\frac{\partial}{\partial\theta}\barphi^b\p^+\barphi^a\frac{\p}{\p^{+2}}\phi^c
&=2\,\theta^i{\bar\theta_i}(f^{abc} \p^+\barphi^b\p^+\barphi^a)\frac{\p}{\p^{+2}}\phi^c=0,
\end{align}
which leads that
\begin{align}
{\rm II+III}&=-2i\sqrt2
f^{abc}\phi^a\frac{\partial}{\partial^+{}^2}\left(\bar\theta\frac{\partial}{\partial\bar\theta}\bar\phi^b\partial^+\bar\phi^c\right).
\end{align}
Hence, 
\begin{align}
{\rm I+II+III}
&= 4i\sqrt2 f^{abc}\frac{1}{\partial^+}\phi^a\bar\phi^b\partial\bar\phi^c 
-2i\sqrt2
f^{abc}\phi^a\frac{\partial}{\partial^+{}^2}\left(\bar\theta\frac{\partial}{\partial\bar\theta}\bar\phi^b\partial^+\bar\phi^c\right),
\end{align}
which should be the same as \eqref{startingpt} and thus yields that
\begin{align}
i\sqrt2
f^{abc}\int  \phi^a \frac{\p}{\partial^{+2}} ({\bar \theta}_i\frac{\p}{\p\bar\theta_i}
{\bar \phi}^b \partial^+{\bar \phi}^c)
&= \frac{4}{3}i\sqrt2 f^{abc} \int \frac{1}{\partial^+}\phi^a\bar\phi^b\partial\bar\phi^c, 
\end{align}
This proves the identity \eqref{identity}.\\

\noindent{\underline{\bf Second 3-point function identity }}\\
Another useful identity is 
\begin{align}
&f^{abc}\int \frac{\p_3}{\p^+}q^i \phi^a C_{ij}\frac{1}{\p^{+2}}(q^j\bar\phi^b\p^+\barphi^c)\cr
&=~\frac{1}{3}f^{abc}\int\left(\frac{1}{\p^+}\phi^a\barphi^b \frac{d^iC_{ij}d^j}{\p^+}{\p_3} \barphi^c+
\frac12 \phi^a\barphi^b \frac{d^iC_{ij}d^j}{\p^{+2}}{\p_3} \barphi^c\right).
\label{3pt-id2}
\end{align}\\
To prove this identity,  
we first use \eqref{qphi} to write the LHS of \eqref{3pt-id2} as
\begin{align}
-i\sqrt2 f^{abc}\int\frac{\p_3}{\p^{+2}}\phi^a \theta^iC_{ij}\frac{\p}{\p\bar\theta_j}\barphi^b {\p^+}\barphi^c.\label{3pt-1}
\end{align}
The integrations by parts with respect to $\frac{\p}{\p\bar\theta_j}$ then yield
\begin{align}
i\sqrt2 f^{abc}\int\theta^iC_{ij}\frac{\p}{\p\bar\theta_j}\frac{\p_3}{\p^{+2}}\phi^a \barphi^b {\p^+}\barphi^c
+
i\sqrt2 f^{abc}\int\frac{\p_3}{\p^{+2}}\phi^a \barphi^b {\p^+}\theta^iC_{ij}\frac{\p}{\p\bar\theta_j}\barphi^c\equiv{\sf IV}+{\sf V}.\label{3pt-2}
\end{align}
Applying the inside-out constraint on $\barphi^c$ followed by integrations by parts with respect to $\bar d's$, 
we find that the first term of \eqref{3pt-2}, ${\sf IV}$, can be rewritten as (dropping the integral symbols and $f^{abc}$)
\begin{align}
{\sf IV}&=\frac{i\sqrt2}{2\cdot 4!}\ep^{ijkl}\Big(\bar d_i\bar d_j\bar d_k\bar d_l
 \theta\frac{\p}{\p\bar\theta}\frac{\p_3}{\p^{+2}}\phi^a
\Big)\barphi^b\frac{1}{\p^+}\barphi^c\cr
&= -\frac{d^iC_{ij} d^j}{2}\frac{\p_3}{\p^+}\barphi^a\barphi^b\frac{1}{\p^+}\phi^c 
+ i\sqrt2 
 \theta\frac{\p}{\p\bar\theta}\p_3\barphi^a\barphi^b\frac{1}{\p^+}\phi^c,\label{3pt-2-res}
\end{align}
where $\theta\frac{\p}{\p\bar\theta}=\theta^iC_{ij}\frac{\p}{\p\bar\theta_j}$.

We now consider the second term of \eqref{3pt-2}, ${\sf V}$. The integration by parts with respect to $\p^+$ acting on $\barphi^c$ yields
\begin{align}
{\sf V}=-i\sqrt2\frac{\p_3}{\p^+}\phi^a\barphi^b\theta\frac{\p}{\p\bar\theta}\barphi^c
-i\sqrt2\frac{\p_3}{\p^{+2}}\phi^a\p^+\barphi^b\theta\frac{\p}{\p\bar\theta}\barphi^c.
\label{3pt-2-1}
\end{align}
Recognize that the last term is the same as \eqref{3pt-1} with the opposite sign and thus this term adds to \eqref{3pt-1}. We then perform the integration by parts with respect to $\p_3$ to rewrite the first term of \eqref{3pt-2-1} as 
\begin{align}
i\sqrt2\frac{1}{\p^+}\phi^a\p_3\barphi^b\theta\frac{\p}{\p\bar\theta}\barphi^c
+
i\sqrt2\frac{1}{\p^+}\phi^a\barphi^b\theta\frac{\p}{\p\bar\theta}\p_3\barphi^c,\label{3pt-2-2}
\end{align}
Using the inside-out constraint on $\barphi^b$ and
\eqref{usefulcomm2}, we find that the first term of \eqref{3pt-2-2} is written as
\begin{align}
i\sqrt2\p^+\barphi^a\theta\frac{\p}{\p\bar\theta}\barphi^c\frac{\p_3}{\p^{+2}}\phi^b
-\frac{1}{2}\phi^a\barphi^c\frac{\p_3}{\p^{+2}}d^kC_{kl}d^l\barphi^b
-\frac{1}{2}\frac{1}{\p^+}\phi^a\barphi^c\frac{\p_3}{\p^{+}}d^kC_{kl}d^l\barphi^b,
\end{align}
where the first term is also of the same form as \eqref{3pt-1} and thus adds to \eqref{3pt-1}. 

Since the second term of \eqref{3pt-2-2} is cancelled by the last term of \eqref{3pt-2-res}, we hence obtain that  
\begin{align}
{\sf IV}+{\sf V}=&
2i\sqrt2 \frac{\p_3}{\p^{+2}}\phi^a \theta^iC_{ij}\frac{\p}{\p\bar\theta_j}\barphi^b {\p^+}\barphi^c
-{d^iC_{ij} d^j}\frac{\p_3}{\p^+}\barphi^a\barphi^b\frac{1}{\p^+}\phi^c\cr
&
-\frac{1}{2}\frac{1}{\p^+}\phi^a\barphi^c\frac{\p_3}{\p^{+}}d^kC_{kl}d^l\barphi^b,
\end{align}
and this leads to the identity \eqref{3pt-id2}.


\setcounter{equation}{0}
\section{Spinors and bilinears in spinor space for $SO(5)$}
For an $SO(5)$, one can choose   
five  $4 \times 4$ $\gamma$-matrices satisfying 
\begin{align}
\{ \gamma^I, \gamma^J \} = 2 \delta^{IJ}, \hspace{1cm} I= 1,2,\ldots 5.
\end{align}
These $\gamma$-matrices are all hermitian. However, they have different symmetry properties under interchange of their spinor indices. We can then find a matrix, $C$, which is antisymmetric and makes 
\begin{align*}
&C\gamma^I&&{\rm antisymmetric} & {\bf 5} \cr
&C\gamma^{IJ}&&{\rm symmetric}&{\bf 10}\cr
&C\gamma^{IJK}= \epsilon^{IJKLM} C\gamma^{LM}&& {\rm symmetric}&\cr
&C\gamma^{IJKL}= \epsilon^{IJKLM} C\gamma^{M}&& {\rm antisymmetric}&\cr
&C\gamma^{IJKLM}= \epsilon^{IJKLM} C&& {\rm antisymmetric}&
\end{align*}
\noindent We take a Lorentz transformation of a $4$-component spinor to be
\begin{align}
\delta \psi = \frac{1}{2} \gamma^{IJ} \psi .
\end{align}
We see easily that we can form Lorentz covariant expression from two spinors as 
\begin{align}
\bar \psi \, \gamma^{I\ldots L} \lambda, && \bar\psi \equiv \psi^{\dagger}.
\end{align}
There is, however, another type of covariant expression we can form 
\begin{align}
\psi \, C\gamma^{I\ldots L} \lambda.
\end{align}
A specific representation of the matrix $C$ is 
\begin{align}
C_{ij}=C^{ij}= \begin{pmatrix}
&&&1\\
&&1&\\
&-1&&\\
-1&&&
\end{pmatrix}.
\end{align}


\providecommand{\href}[2]{#2}\begingroup\raggedright\endgroup

\end{document}